\documentclass[prl,showpacs,twocolumn,superscriptaddress,amsmath,amssymb]{revtex4}

\usepackage{multirow}
\usepackage{graphicx}
\newcommand\ls{\hspace{0pt}} 
\newcommand\lss{\hspace{5pt}} 
\newcommand{\dof}{DOF}
\newcommand{\oam}{OAM}
\newcommand{\vio}[2]{\ensuremath{#1[#2\sigma]}}

\begin{document}

\title{Generation of Hyperentangled Photon Pairs} 
\author{Julio T. Barreiro} \affiliation{Department of Physics, University of
  Illinois at Urbana-Champaign, Urbana, IL 61801-3080, USA} 
\author{Nathan K. Langford} \affiliation{Department of Physics, University of
  Queensland, Brisbane, QLD 4072, Australia}
\author{Nicholas A. Peters} \affiliation{Department of Physics, University of
  Illinois at Urbana-Champaign, Urbana, IL 61801-3080, USA} 
\author{Paul G. Kwiat} \affiliation{Department of Physics, University of
  Illinois at Urbana-Champaign, Urbana, IL 61801-3080, USA} 
\date{\today} 

\begin{abstract}
We experimentally demonstrate the first quantum system entangled in
\emph{every} degree of freedom (hyperentangled).  Using pairs of photons
produced in spontaneous parametric down-conversion, we verify entanglement by
observing a Bell-type inequality violation in each degree of freedom:
polarization, spatial mode and time-energy.  We also produce and characterize
maximally hyperentangled states and novel states simultaneously exhibiting both
quantum and classical correlations.  Finally, we report the tomography of a
2x2x3x3 system (36-dimensional Hilbert space), which we believe is the first
reported photonic entangled system of this size to be so characterized.
\end{abstract}

\maketitle 

Entanglement, the quintessential quantum mechanical correlations that can
exist between quantum systems, plays a critical role in many important
applications in quantum information processing, including the revolutionary
one-way quantum computer~\cite{raussendorf:prl-86-5188}, quantum
cryptography~\cite{ekert:prl-67-661}, dense coding~\cite{bennett:prl-69-2881}
and teleportation~\cite{bennett:pr-70-1895}.  As a result, the ability to
create, control and manipulate entanglement has been a defining experimental
goal in recent years.  Higher-order entanglement has been realized in
multi-particle~\cite{zhao:nat-430-54} and
multi-dimensional~\cite{mair:nature-412-313,thew:prl-93-010503,osullivan-hale:prl-94-220501,oemrawsingh:quant-ph-0506253}
systems.  Furthermore, two-component quantum systems can be entangled in
every degree of freedom (\dof{}), or
hyperentangled~\cite{kwiat:jmo-44-2173}.  These systems enable the
implementation of 100\%-efficient complete Bell-state analysis with only
linear elements~\cite{kwiat:pra-58-R2623} and techniques for state
purification~\cite{simon:prl-89-257901}.  In addition, hyperentanglement can
also be interpreted as entanglement between two higher-dimensional quantum
systems, offering significant advantages in quantum communication protocols
(e.g., secure superdense coding~\cite{wang:pra-71-044305} and
cryptography~\cite{bruss:prl-88-127901}).

Photon pairs produced via the nonlinear optical process of spontaneous
parametric down-conversion have many accessible \dof{} which can be exploited
for the production of entanglement.  This was first demonstrated using
polarization~\cite{ou:prl-61-50,shih:prl-61-2921}, but the list expanded
rapidly to include momentum (linear~\cite{rarity:prl-64-2495},
orbital~\cite{mair:nature-412-313}, and
transverse~\cite{langford:prl-93-053601} spatial modes),
energy-time~\cite{franson:prl-62-2205} and
time-bin~\cite{brendel:prl-82-2594}, simultaneous polarization and
energy-time~\cite{strekalov:pra-54-R1}, and recently, simultaneous
polarization and 2-level linear momentum~\cite{yang:quant-ph-0502085}.
In this work, we produce and characterize pairs of single photons
simultaneously entangled in \emph{every} \dof{} --polarization, spatial mode
and energy-time.  As observed previously~\cite{mair:nature-412-313}, photon
pairs from a \emph{single} nonlinear crystal are entangled in orbital angular
momentum (\oam{}).  Moreover, polarization entangled states can be created by
coherently pumping \emph{two} contiguous thin
crystals~\cite{kwiat:pra-60-R773}, provided the spatial modes emitted from
each crystal are indistinguishable.  Finally, the pump distributes energy to
the daughter photons in many ways, entangling each pair in energy;
equivalently, each pair is coherently emitted over a range of times (within
the coherence of the continuous wave pump).  We show our two-crystal source
can generate a $2\times2\times3\times3\times2\times2$-dimensional
hyperentangled state~\cite{kwiat:jmo-44-2173}, approximately
\begin{equation}
\underbrace{\left(|HH\rangle+|VV\rangle\right)}_\text{polarization}\otimes
\underbrace{\left(|rl\rangle + \alpha|gg\rangle+|lr\rangle\right)}_\text{spatial modes}\otimes
\underbrace{\left(|ss\rangle+|ff\rangle\right)}_\text{energy-time}\,.\label{eq:hyper}
\end{equation}
Here $H$ ($V$) represents the horizontal (vertical) photon polarization;
$|l\rangle$, $|g\rangle$ and $|r\rangle$, represent the paraxial spatial
modes (Laguerre-Gauss) carrying $-\hbar$, 0, and $+\hbar$ \oam{},
respectively~\cite{allen:2003-oam}; $\alpha$ describes the \oam{} spatial
mode balance prescribed by the source~\cite{molina:prl-88-013601} and
selected via the mode-matching conditions; and $|s\rangle$ and $|f\rangle$,
respectively, represent the relative early and late emission times of a pair
of energy anticorrelated photons~\cite{franson:prl-62-2205}.

The most common maximally entangled states are the 2-qubit Bell states:
$\Phi^\pm=(|00\rangle\pm|11\rangle)/\sqrt{2}$ and
$\Psi^\pm=(|01\rangle\pm|10\rangle)/\sqrt{2}$, in the logical basis
$|0\rangle$ and $|1\rangle$.  By collecting only the $\pm\hbar$ \oam{} state
of the spatial subspace, the state (\ref{eq:hyper}) becomes a tensor product
of three Bell states
$\Phi^+_\text{poln}\otimes\Phi^+_\text{spa}\otimes\Phi^+_\text{t-e}$.  As a
preliminary test of the hyperentanglement, we characterized the polarization
and spatial mode subspaces by measuring the entanglement (characterized by
tangle $T$~\cite{wootters:prl-80-2245}), the mixture (characterized by linear
entropy
$S_L(\rho)=\frac{4}{3}[1-\text{Tr}(\rho^2)]$~\cite{james:pra-64-052312}), and
the fidelity $F(\rho,\rho_t) \equiv
(\text{Tr}(\sqrt{\sqrt{\rho_t}\rho\sqrt{\rho_t}}))^2$ of the measured state
$\rho$ with the target state $\rho_t=|\psi_t\rangle\langle\psi_t|$.  We
consistently measured high-quality states with tangles, linear entropies, and
fidelities with $\Phi^+$ of $T=0.99(1)$, $S_L=0.01(1)$ and $F=0.99(1)$ for
polarization; and $T=0.96(1)$, $S_L=0.03(1)$ and $F=0.95(1)$ for spatial
mode, significantly higher than earlier
results~\cite{langford:prl-93-053601}.
\begin{figure}[b]
\centerline{\includegraphics{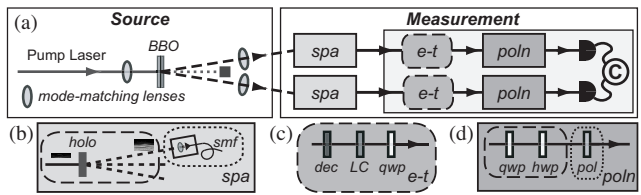}}
\caption{Experimental setup for the creation and analysis of hyperentangled
  photons.
(a) The photons, produced using adjacent nonlinear crystals (BBO),
  pass through a state filtration process for each \dof{} before
  coincidence detection.  The measurement insets show the filtration
  processes as a transformation of the target state (dashed box) and a
  filtering step to discard the other components of the state (dotted box).
(b) \emph{Spatial Filtration (spa)}: hologram (holo) and single-mode
  fiber (smf). 
(c) \emph{Energy-Time Transformation (e-t)}: thick quartz decoherer
  (dec) and liquid crystal (LC).
(d) \emph{Polarization Filtration (poln)}: quarter-wave plate (qwp),
  half-wave plate (hwp) and polarizer (pol).
}\label{fig:setup}
\end{figure}
The experiment is illustrated in Fig.~1.  A 120-mW 351-nm Ar$^+$ laser pumps
two contiguous $\beta$-barium borate (BBO) nonlinear crystals with optic axes
aligned in perpendicular planes~\cite{kwiat:pra-60-R773}.  Each 0.6-mm thick
crystal is phasematched to produce Type-I degenerate photons at 702 nm into a
cone of 3.0$^\circ$ half-opening angle.  The first (second) crystal produce
pairs of horizontally (vertically) polarized photons and these two possible
down-conversion processes are coherent, provided the spatial modes emitted
from each crystal are indistinguishable.  With the pump focused to a waist at
the crystals, this constraint can be satisfied by using thin crystals and
``large'' beam waists (large relative to the mismatch in the overlap of the
down-conversion cones from each crystal~\cite{kwiat:pra-60-R773}). However,
the \oam{}-entanglement is maximized by balancing the relative populations of
the low-valued \oam{} eigenstates~\cite{molina:prl-88-013601}, which requires
smaller beam waists to image a large area of the down-conversion cones.  Here
we compromise by employing an intermediate waist size ($\sim\!90\,\mu$m) at
the crystal.  Mode-matching lenses are then used to optimize the coupling of
the rapidly diverging down-conversion modes into single-mode collection
fibers.

The measurement process consists of three stages of local state projection,
one for each \dof{}.  At each stage, the target state is transformed into a
state that can be discriminated from the other states with high accuracy.
Specifically, computer-generated phase holograms~\cite{binary} transform the
target spatial mode into the pure gaussian (or 0-\oam{}) mode, which is then
filtered by the single-mode fiber~\cite{mair:nature-412-313} (Fig.~1b).
After a polarization controller to compensate for the fiber birefringence,
wave plates transform the target polarization state into horizontal, which is
filtered by a polarizer (Fig.~1d).  The analysis of the energy-time state is
realized by a Franson-type~\cite{franson:prl-62-2205} polarization
interferometer without detection post-selection~\cite{strekalov:pra-54-R1}.
The matched unbalanced interferometers give each photon a fast $|f\rangle$
and slow $|s\rangle$ route to its detector.  Our interferometers consisted of
$L\!\sim\!11$-mm quartz birefringent elements, which longitudinally separated
the horizontal and vertical polarization components by
$\Delta{}n_\text{quartz}L\!\sim\!100\,\mu$m, more than the single-photon
coherence length ($\lambda^2/\Delta\lambda\!\sim\!50\;\mu$m with
$\Delta\lambda\!=\!10\,$nm from the interference filters) but much less than
the pump-photon coherence length ($\sim10\,$cm).  We rely on the photons'
polarization entanglement $|HH\rangle+|VV\rangle$ to thus project onto a
two-time state ($|Hs,Hs\rangle+e^{i(\delta_1+\delta_2)} |Vf,Vf\rangle$),
where $\delta_1$ and $\delta_2$ are controlled by birefringent elements
(liquid crystals and quarter-wave plates) in the path of each
photon~\cite{strekalov:pra-54-R1}.  Finally, by analyzing the polarization in
the $\pm45^\circ$ basis, we erase the distinguishing polarization labels and
can directly measure the coherence between the $|ss\rangle$ and $|ff\rangle$
terms, arising from the energy-time entanglement.

To verify quantum mechanical correlations, we tested every \dof{} against a
Clauser-Horne-Shimony-Holt (CHSH) Bell inequality~\cite{clauser:prl-23-880}.
The CHSH inequality places constraints ($S\le2$) on the value of the Bell
parameter $S$, a combination of four two-particle correlation probabilities
using two possible analysis settings for each photon.  If $S>2$, no separable
quantum system (or local hidden variable theory) can explain the
correlations; in this sense, a Bell inequality acts as an ``entanglement
witness''~\cite{terhal:pla-271-319}.  To measure the strongest violation for
the polarization and spatial-mode \dof{}s, we determined the optimal
measurement settings by first tomographically reconstructing the 2-qubit
subspace of interest; we employ a maximum likelihood technique to identify
the density matrix most consistent with the data~\cite{james:pra-64-052312}.

\begin{table}[b]
\caption{Bell parameter $S$ showing CHSH-Bell inequality violations in every
    degree of freedom.  The local realistic limit ($S\le2$) is violated by
    the number of standard deviations shown in brackets, determined by
    counting statistics.}
\begin{ruledtabular}
\begin{tabular}{c@{\lss}c@{\ls}c@{\ls}c@{\ls}c@{\ls}c}
\multirow{2}{*}{\dof{}}& \multicolumn{5}{c}{Spatial mode projected subspaces} \\
& $|gg\rangle\langle gg|$ 
& $|rl\rangle\langle rl|$ 
& $|lr\rangle\langle lr|$ 
& $|hh\rangle\langle hh|$ 
& $|vv\rangle\langle vv|$ 
\\\hline
$\Phi^+_\text{poln}$
& \vio{2.76}{76}
& \vio{2.78}{46} 
& \vio{2.75}{44} 
& \vio{2.81}{40}
& \vio{2.75}{33}\\
$\Phi^+_\text{t-e}$
& \vio{2.78}{77} 
& \vio{2.80}{40} 
& \vio{2.80}{40}
& \vio{2.72}{30}
& \vio{2.74}{29}\\\hline\hline
\multicolumn{2}{c}{\multirow{2}{*}{\dof{}}} &\multicolumn{3}{c}{Polarization projected subspaces}\\
&& no polarizers & $|HH\rangle\langle HH|$ & $|VV\rangle\langle VV|$ \\\hline
\multicolumn{2}{c}{$\Phi^+_\text{spa}$}
& \vio{2.78}{78} 
& \vio{2.80}{36}
& \vio{2.79}{37}\\
\multicolumn{2}{c}{$\alpha|gg\rangle + |rl\rangle$}
& \vio{2.33}{55} 
& \vio{2.30}{25}
& \vio{2.38}{30} \\
\multicolumn{2}{c}{$\alpha|gg\rangle + |lr\rangle$}
& \vio{2.28}{47} 
& \vio{2.26}{20}
& \vio{2.31}{26}
\end{tabular}
\end{ruledtabular}
\label{table:vio}
\vspace{-5mm}
\end{table}
Table~\ref{table:vio} shows the Bell parameters measured for the
polarization, spatial mode, and energy-time subspaces, with various
projections in the complementary \dof{}.  We see that for every subspace, the
Bell parameter exceeded the classical limit of $S=2$ by more than 20 standard
deviations ($\sigma$), verifying the hyperentanglement.  For both the
polarization and spatial-mode measurements, we traced over the energy-time
\dof{} by not projecting in this subspace.  We measured the polarization
correlations while projecting the spatial modes into the orthogonal basis
states ($|l\rangle,|g\rangle$, and $|r\rangle$), as well as the
superpositions $|h\rangle\equiv(|l\rangle+|r\rangle)/\sqrt{2}$ and
$|v\rangle\equiv(|l\rangle-|r\rangle)/\sqrt{2}$).  The measured Bell
parameters agreed (within $\sim\!2\,\sigma$) with predictions from
tomographic reconstruction and violated the inequality by more than
$30\,\sigma$.  In the spatial mode \dof{}, the correlations for the state
$\Phi^+_\text{spa}$ were close to maximal ($S=2\sqrt{2}\approx 2.83$), also
in agreement with predictions from the measured state density matrix.  In
addition, we tested Bell inequalities for non-maximally entangled states in
the \oam{}-subspace: $\alpha|gg\rangle + |rl\rangle$ and $\alpha|gg\rangle +
|lr\rangle$; the measured Bell parameters in this case were slightly smaller
(5\%, max.)  than predictions from tomographic reconstruction~\cite{leakage},
yet still $20\,\sigma$ above the classical limit.  Finally, our measured Bell
violation for the energy-time \dof{} using particular phase settings is in
good agreement with the prediction ($S=2\sqrt{2}V$) from the measured
2-photon interference visibility $V=0.985(2)$.

\begin{figure}[b]
\centerline{\includegraphics{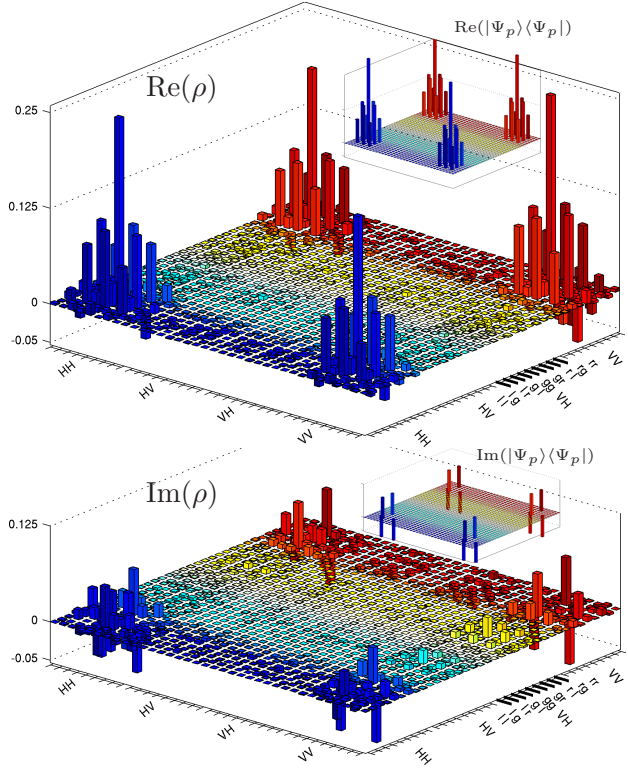}}
\caption{(color online). Measured density matrix ($\rho$) and close pure state
  ($|\Psi_p\rangle\sim\Phi^+_\text{poln}\otimes(|lr\rangle+\alpha|gg\rangle+|rl\rangle)$
  with $\alpha=1.88e^{0.16i\pi}$) of a
  (2$\times$2$\times$3$\times$3)-dimensional state of 2-photon polarization and
  spatial mode~\cite{tomodetails}.}
\label{fig:thebig}
\end{figure}
\begin{figure}
\centerline{\includegraphics{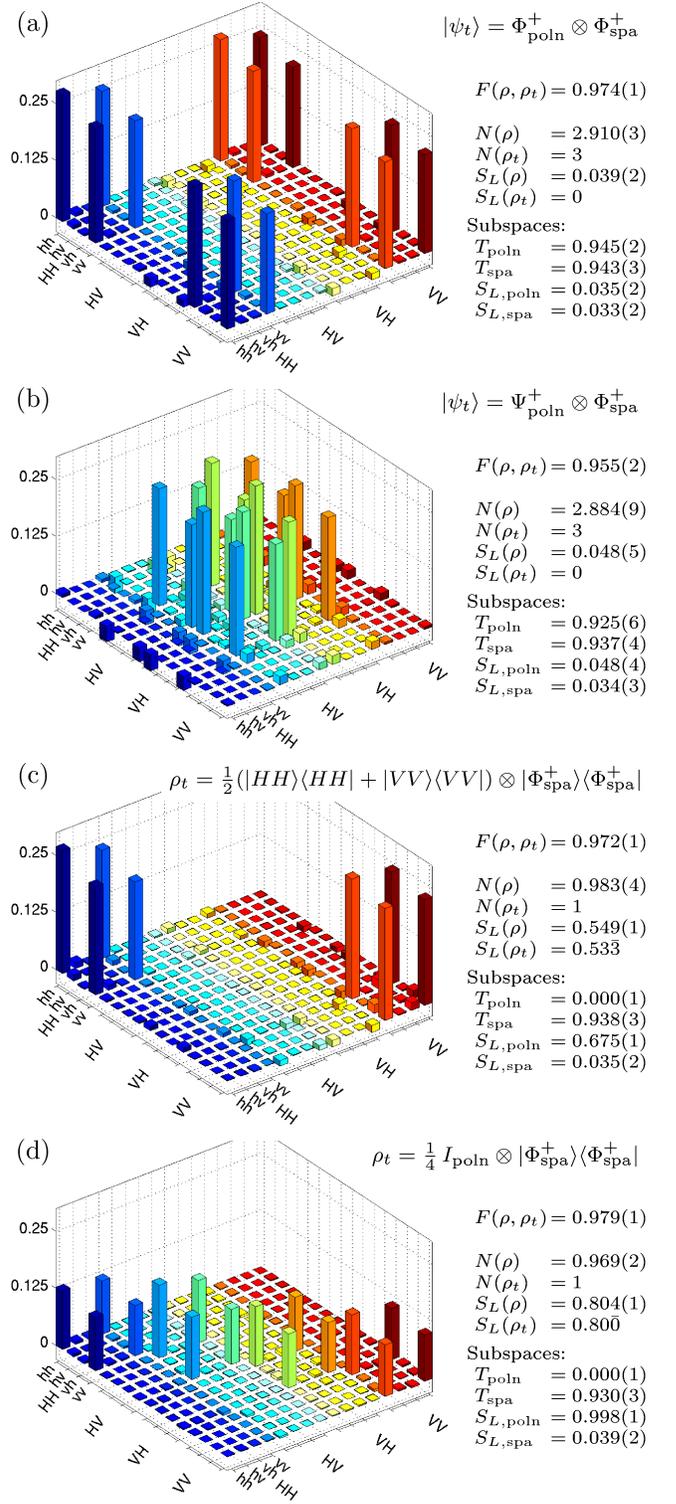}}
\caption{(color online). Measured density matrices (real parts) of
  (2$\times$2$\times$2$\times$2)-dimensional states of 2-photon polarization
  and ($+1,-1$)-qubit \oam{}~\cite{tomodetails}.  For each state, we list: the
  target state $\rho_t$, the fidelity $F(\rho,\rho_t)$ of the measured state
  $\rho$ with the target $\rho_t$, their negativities and linear entropies, and
  the tangle and linear entropy for each subspace.  The negativity for
  two-qubit states is the square root of the tangle.  The magnitudes of all
  imaginary elements, not shown, are less than 0.03.}
\label{fig:results}
\end{figure}
The polarization and spatial-mode state was fully characterized via
tomography~\cite{james:pra-64-052312}.  We performed the 1296 linearly
independent state projections required for a full reconstruction in the
$(2\otimes3)\otimes(2\otimes3)$ Hilbert space consisting of two polarization
and three \oam{} modes for each photon.  The measured state (Fig.~2) overlaps
the anticipated state (polarization and spatial \dof{}s of Eq. 1) with a
fidelity of $0.69(1)$ for $\alpha\!=\!1.88e^{0.16i\pi}$ (numerically fitted),
and $S_L\!=\!0.46(1)$, suggesting the difference arises mostly from mixture.
Treating the photon pairs as a six-level two-particle system, we can quantify
the entanglement using the negativity $N$~\cite{zyczkowski:pra-58-883}.  In
this $6\otimes6$ Hilbert space, $N$ ranges from 0 (for separable states) to 5
(for maximally-entangled states), and the fitted state above has
$N\!\approx4.44\!$.  Our measured partially mixed state has $N\!=\!2.96(4)$,
indicating strong entanglement.  The spatial mode alone has $N\!=\!1.14(2)$,
greater than the maximum ($N\!=\!1$) of any two-qubit system.  Thus, our
large state possesses 2-qubit and 2-qutrit entanglement.

We also selected a state (neglecting the $|gg\rangle$ component, Fig.~3a)
maximally entangled in both polarization and spatial mode, that had
$F=0.974(1)$ with the target $\Phi^+_\text{poln}\otimes\Phi^+_\text{spa}$.
By tracing over polarization (spatial mode), we look at the measured state in
the spatial mode (polarization) subspaces.  The reduced states in both
\dof{}s are pure ($S_L\!<\!0.04$) and highly entangled ($T\!>\!0.94$).

With this precise source of hyperentanglement, we have the flexibility to
prepare nearly arbitrary polarization states~\cite{white:pra-65-012301}, and
to select arbitrary spatial-mode encodings.  For example, we also generated a
different maximally entangled state:
$\Psi^+_\text{poln}\otimes\Phi^+_\text{spa}$ (Fig.~3b).  By coupling to and
tracing over the energy-time \dof{} using quartz
decoherers~\cite{white:pra-65-012301}, we can add mixture to the polarization
subspace, allowing us to prepare a previously unrealized state that
simultaneously displays \emph{classical correlations} in polarization and
maximal \emph{quantum correlations} between spatial modes~(Fig.~3c):
$\rho\approx\frac{1}{2}(|HH\rangle\langle{}HH|+|VV\rangle\langle{}VV|)\otimes|\Phi^+_\text{spa}\rangle\langle\Phi^+_\text{spa}|$.
We were also able to accurately prepare the state
$\rho_t=\frac{1}{4}I_\text{poln}\otimes|\Phi^+_\text{spa}\rangle\langle\Phi^+_\text{spa}|$,
with no polarization correlations at all (i.e., completely mixed or
unpolarized), while still maintaining near maximal entanglement in the
spatial \dof{} (Fig.~3d).

We report the first realization of hyperentanglement of a pair of single
photons.  The entanglement in each \dof{} is demonstrated by violations of
CHSH-Bell inequalities of greater than $20\sigma$.  Also, using tomography we
fully characterize a $2\!\otimes\!2\!\otimes\!3\!\otimes\!3$ state, the
largest quantum system to date.  In restricted
($2\!\times\!2\!\times\!2\!\times\!2$)-dimensional subspace, we
prepare a range of target states with unprecedented fidelities for quantum
systems of this size, including novel states with a controllable degree of
correlation in the polarization subspace.  These hyperentangled states
enable 100\%-efficient Bell-state analysis~\cite{kwiat:pra-58-R2623}, which
is important for a variety of quantum information
protocols~\cite{bennett:prl-69-2881,wang:pra-71-044305}.  Because the spatial
mode and energy-time \dof{}s are infinite in size, we envision examining even
larger subspaces, encoding higher-dimensional
qudits~\cite{thew:prl-93-010503,osullivan-hale:prl-94-220501}.  Finally, we
note that the pairwise mechanism of the $\chi^{(2)}$ down-conversion process
inherently produces entanglement in photon
number~\cite{eisenberg:prl-93-193901}.

We thank A. G. White and T.-C. Wei for helpful discussions and the
ARO/ARDA-sponsored MURI Center for Photonic Quantum Information Systems, the
ARC, and the U. Queensland Found. for support.  J.T.B. acknowledges support
from CONACYT-M\'{e}xico.


\begin{thebibliography}{32}
\expandafter\ifx\csname natexlab\endcsname\relax\def\natexlab#1{#1}\fi
\expandafter\ifx\csname bibnamefont\endcsname\relax
  \def\bibnamefont#1{#1}\fi
\expandafter\ifx\csname bibfnamefont\endcsname\relax
  \def\bibfnamefont#1{#1}\fi
\expandafter\ifx\csname citenamefont\endcsname\relax
  \def\citenamefont#1{#1}\fi
\expandafter\ifx\csname url\endcsname\relax
  \def\url#1{\texttt{#1}}\fi
\expandafter\ifx\csname urlprefix\endcsname\relax\def\urlprefix{URL }\fi
\providecommand{\bibinfo}[2]{#2}
\providecommand{\eprint}[2][]{\url{#2}}

\bibitem[{\citenamefont{Raussendorf and
  Briegel}(2001)}]{raussendorf:prl-86-5188}
\bibinfo{author}{\bibfnamefont{R.}~\bibnamefont{Raussendorf}} \bibnamefont{and}
  \bibinfo{author}{\bibfnamefont{H.~J.} \bibnamefont{Briegel}},
  \bibinfo{journal}{Phys. Rev. Lett.} \textbf{\bibinfo{volume}{86}},
  \bibinfo{pages}{5188} (\bibinfo{year}{2001}).

\bibitem[{\citenamefont{Ekert}(1991)}]{ekert:prl-67-661}
\bibinfo{author}{\bibfnamefont{A.~K.} \bibnamefont{Ekert}},
  \bibinfo{journal}{Phys. Rev. Lett.} \textbf{\bibinfo{volume}{67}},
  \bibinfo{pages}{661} (\bibinfo{year}{1991}).

\bibitem[{\citenamefont{Bennett and Wiesner}(1992)}]{bennett:prl-69-2881}
\bibinfo{author}{\bibfnamefont{C.~H.} \bibnamefont{Bennett}} \bibnamefont{and}
  \bibinfo{author}{\bibfnamefont{S.~J.} \bibnamefont{Wiesner}},
  \bibinfo{journal}{Phys. Rev. Lett.} \textbf{\bibinfo{volume}{69}},
  \bibinfo{pages}{2881} (\bibinfo{year}{1992}).

\bibitem[{\citenamefont{Bennett et~al.}(1993)}]{bennett:pr-70-1895}
\bibinfo{author}{\bibfnamefont{C.~H.} \bibnamefont{Bennett}}
  \bibnamefont{et~al.}, \bibinfo{journal}{Phys. Rev. Lett.}
  \textbf{\bibinfo{volume}{70}}, \bibinfo{pages}{1895} (\bibinfo{year}{1993}).

\bibitem[{\citenamefont{Zhao et~al.}(2004)}]{zhao:nat-430-54}
\bibinfo{author}{\bibfnamefont{Z.}~\bibnamefont{Zhao}} \bibnamefont{et~al.},
  \bibinfo{journal}{Nature} \textbf{\bibinfo{volume}{430}}, \bibinfo{pages}{54}
  (\bibinfo{year}{2004}); 
\bibinfo{author}{\bibfnamefont{P.}~\bibnamefont{Walther}} \bibnamefont{et~al.},
  \bibinfo{journal}{Nature} \textbf{\bibinfo{volume}{434}},
  \bibinfo{pages}{169} (\bibinfo{year}{2005}{\natexlab{a}}).

\bibitem[{\citenamefont{Mair et~al.}(2001)\citenamefont{Mair, Vaziri, Weihs,
  and Zeilinger}}]{mair:nature-412-313}
\bibinfo{author}{\bibfnamefont{A.}~\bibnamefont{Mair}},
  \bibinfo{author}{\bibfnamefont{A.}~\bibnamefont{Vaziri}},
  \bibinfo{author}{\bibfnamefont{G.}~\bibnamefont{Weihs}}, \bibnamefont{and}
  \bibinfo{author}{\bibfnamefont{A.}~\bibnamefont{Zeilinger}},
  \bibinfo{journal}{Nature} \textbf{\bibinfo{volume}{412}},
  \bibinfo{pages}{313} (\bibinfo{year}{2001}).

\bibitem[{\citenamefont{Thew et~al.}(2004)\citenamefont{Thew, Ac\'{i}n,
  Zbinden, and Gisin}}]{thew:prl-93-010503}
\bibinfo{author}{\bibfnamefont{R.~T.} \bibnamefont{Thew}},
  \bibinfo{author}{\bibfnamefont{A.}~\bibnamefont{Ac\'{i}n}},
  \bibinfo{author}{\bibfnamefont{H.}~\bibnamefont{Zbinden}}, \bibnamefont{and}
  \bibinfo{author}{\bibfnamefont{N.}~\bibnamefont{Gisin}},
  \bibinfo{journal}{Phys. Rev. Lett.} \textbf{\bibinfo{volume}{93}},
  \bibinfo{pages}{10503} (\bibinfo{year}{2004}).

\bibitem[{\citenamefont{OSullivan-Hale
  et~al.}(2005)\citenamefont{OSullivan-Hale, Khan, Boyd, and
  Howell}}]{osullivan-hale:prl-94-220501}
\bibinfo{author}{\bibfnamefont{M.~N.} \bibnamefont{OSullivan-Hale}},
  \bibinfo{author}{\bibfnamefont{I.~A.} \bibnamefont{Khan}},
  \bibinfo{author}{\bibfnamefont{R.~W.} \bibnamefont{Boyd}}, \bibnamefont{and}
  \bibinfo{author}{\bibfnamefont{J.~C.} \bibnamefont{Howell}},
  \bibinfo{journal}{Phys. Rev. Lett.} \textbf{\bibinfo{volume}{94}},
  \bibinfo{pages}{220501} (\bibinfo{year}{2005}).

\bibitem[{\citenamefont{Oemrawsingh et~al.}()}]{oemrawsingh:quant-ph-0506253}
  \bibinfo{author}{\bibfnamefont{S.}~\bibnamefont{Oemrawsingh}}
  \bibnamefont{et~al.}, \bibinfo{journal}{Phys. Rev. Lett.}
  \textbf{\bibinfo{volume}{95}}, \bibinfo{pages}{240501}
  (\bibinfo{year}{2005}).

\bibitem[{\citenamefont{Kwiat}(1997)}]{kwiat:jmo-44-2173}
\bibinfo{author}{\bibfnamefont{P.~G.} \bibnamefont{Kwiat}},
  \bibinfo{journal}{J. Mod. Opt.} \textbf{\bibinfo{volume}{44}},
  \bibinfo{pages}{2173} (\bibinfo{year}{1997}).

\bibitem[{\citenamefont{Kwiat and Weinfurter}(1998)}]{kwiat:pra-58-R2623}
\bibinfo{author}{\bibfnamefont{P.~G.} \bibnamefont{Kwiat}} \bibnamefont{and}
  \bibinfo{author}{\bibfnamefont{H.}~\bibnamefont{Weinfurter}},
  \bibinfo{journal}{Phys. Rev. A} \textbf{\bibinfo{volume}{58}},
  \bibinfo{pages}{R2623} (\bibinfo{year}{1998}).

\bibitem[{\citenamefont{Simon and Pan}(2002)}]{simon:prl-89-257901}
\bibinfo{author}{\bibfnamefont{C.} \bibnamefont{Simon}} \bibnamefont{and}
  \bibinfo{author}{\bibfnamefont{J.-W.}~\bibnamefont{Pan}},
  \bibinfo{journal}{Phys. Rev. Lett.} \textbf{\bibinfo{volume}{89}},
  \bibinfo{pages}{257901} (\bibinfo{year}{2002}).

\bibitem[{\citenamefont{Wang et~al.}(2005)}]{wang:pra-71-044305}
\bibinfo{author}{\bibfnamefont{C.}~\bibnamefont{Wang}} \bibnamefont{et~al.},
  \bibinfo{journal}{Phys. Rev. A} \textbf{\bibinfo{volume}{71}},
  \bibinfo{pages}{044305} (\bibinfo{year}{2005}).

\bibitem[{\citenamefont{Bruss and Macchiavello}(2002)}]{bruss:prl-88-127901}
\bibinfo{author}{\bibfnamefont{D.}~\bibnamefont{Bruss}} \bibnamefont{and}
  \bibinfo{author}{\bibfnamefont{C.}~\bibnamefont{Macchiavello}},
  \bibinfo{journal}{Phys. Rev. Lett.} \textbf{\bibinfo{volume}{88}},
  \bibinfo{pages}{127901} (\bibinfo{year}{2002});
\bibinfo{author}{\bibfnamefont{N.~J.} \bibnamefont{Cerf}},
  \bibinfo{author}{\bibfnamefont{M.}~\bibnamefont{Bourennane}},
  \bibinfo{author}{\bibfnamefont{A.}~\bibnamefont{Karlsson}}, \bibnamefont{and}
  \bibinfo{author}{\bibfnamefont{N.}~\bibnamefont{Gisin}},
  \bibinfo{journal}{Phys. Rev. Lett.} \textbf{\bibinfo{volume}{88}},
  \bibinfo{pages}{127902} (\bibinfo{year}{2002}).

\bibitem[{\citenamefont{Ou and Mandel}(1988)}]{ou:prl-61-50}
\bibinfo{author}{\bibfnamefont{Z.~Y.} \bibnamefont{Ou}} \bibnamefont{and}
  \bibinfo{author}{\bibfnamefont{L.}~\bibnamefont{Mandel}},
  \bibinfo{journal}{Phys. Rev. Lett.} \textbf{\bibinfo{volume}{61}},
  \bibinfo{pages}{50} (\bibinfo{year}{1988}).

\bibitem[{\citenamefont{Shih and Alley}(1988)}]{shih:prl-61-2921}
\bibinfo{author}{\bibfnamefont{Y.~H.} \bibnamefont{Shih}} \bibnamefont{and}
  \bibinfo{author}{\bibfnamefont{C.~O.} \bibnamefont{Alley}},
  \bibinfo{journal}{Phys. Rev. Lett.} \textbf{\bibinfo{volume}{61}},
  \bibinfo{pages}{2921} (\bibinfo{year}{1988}).

\bibitem[{\citenamefont{Rarity and Tapster}(1990)}]{rarity:prl-64-2495}
\bibinfo{author}{\bibfnamefont{J.~G.} \bibnamefont{Rarity}} \bibnamefont{and}
  \bibinfo{author}{\bibfnamefont{P.~R.} \bibnamefont{Tapster}},
  \bibinfo{journal}{Phys. Rev. Lett.} \textbf{\bibinfo{volume}{64}},
  \bibinfo{pages}{2495} (\bibinfo{year}{1990}).

\bibitem[{\citenamefont{Langford et~al.}(2004)}]{langford:prl-93-053601}
\bibinfo{author}{\bibfnamefont{N.}~\bibnamefont{Langford}}
  \bibnamefont{et~al.}, \bibinfo{journal}{Phys. Rev. Lett.}
  \textbf{\bibinfo{volume}{93}}, \bibinfo{pages}{53601} (\bibinfo{year}{2004}).

\bibitem[{\citenamefont{Franson}(1989)}]{franson:prl-62-2205}
\bibinfo{author}{\bibfnamefont{J.~D.} \bibnamefont{Franson}},
  \bibinfo{journal}{Phys. Rev. Lett.} \textbf{\bibinfo{volume}{62}},
  \bibinfo{pages}{2205} (\bibinfo{year}{1989}).

\bibitem[{\citenamefont{Brendel et~al.}(1999)\citenamefont{Brendel, Gisin,
  Tittel, and Zbinden}}]{brendel:prl-82-2594}
\bibinfo{author}{\bibfnamefont{J.}~\bibnamefont{Brendel}},
  \bibinfo{author}{\bibfnamefont{N.}~\bibnamefont{Gisin}},
  \bibinfo{author}{\bibfnamefont{W.}~\bibnamefont{Tittel}}, \bibnamefont{and}
  \bibinfo{author}{\bibfnamefont{H.}~\bibnamefont{Zbinden}},
  \bibinfo{journal}{Phys. Rev. Lett.} \textbf{\bibinfo{volume}{82}},
  \bibinfo{pages}{2594} (\bibinfo{year}{1999}).

\bibitem[{\citenamefont{Strekalov et~al.}(1996)}]{strekalov:pra-54-R1}
\bibinfo{author}{\bibfnamefont{D.~V.} \bibnamefont{Strekalov}}
  \bibnamefont{et~al.}, \bibinfo{journal}{Phys. Rev. A}
  \textbf{\bibinfo{volume}{54}}, \bibinfo{pages}{R1} (\bibinfo{year}{1996}).

\bibitem[{\citenamefont{T. Yang et~al.}()}]{yang:quant-ph-0502085}
\bibinfo{author}{\bibfnamefont{T.}~\bibnamefont{Yang}}
  \bibnamefont{et~al.}, \bibinfo{journal}{Phys. Rev. Lett.}
  \textbf{\bibinfo{volume}{95}}, \bibinfo{pages}{240406} (\bibinfo{year}{2005});
\bibinfo{author}{\bibfnamefont{C.}~\bibnamefont{Cinelli}}
  \bibnamefont{et~al.}, \bibinfo{journal}{Phys. Rev. Lett.}
  \textbf{\bibinfo{volume}{95}}, \bibinfo{pages}{240405} (\bibinfo{year}{2005}).

\bibitem[{\citenamefont{Kwiat et~al.}(1999)}]{kwiat:pra-60-R773}
\bibinfo{author}{\bibfnamefont{P.~G.} \bibnamefont{Kwiat}}
  \bibnamefont{et~al.}, \bibinfo{journal}{Phys. Rev. A}
  \textbf{\bibinfo{volume}{60}}, \bibinfo{pages}{R773} (\bibinfo{year}{1999}).

\bibitem[{\citenamefont{Allen et~al.}(2003)\citenamefont{Allen, Barnett, and
  Padgett}}]{allen:2003-oam}
\bibinfo{editor}{\bibfnamefont{L.}~\bibnamefont{Allen}},
  \bibinfo{editor}{\bibfnamefont{S.~M.} \bibnamefont{Barnett}},
  \bibnamefont{and} \bibinfo{editor}{\bibfnamefont{M.~J.}
  \bibnamefont{Padgett}}, eds., \emph{\bibinfo{title}{Optical Angular
  Momentum}} (\bibinfo{publisher}{IoP Publishing}, \bibinfo{address}{Bristol},
  \bibinfo{year}{2003}).

\bibitem[{\citenamefont{Torres et~al.}(2002)\citenamefont{Torres, Alexandrescu,
      and Torner}}]{molina:prl-88-013601} \bibinfo{author}{\bibfnamefont{J.~P.}
  \bibnamefont{Torres}},
  \bibinfo{author}{\bibfnamefont{A.}~\bibnamefont{Alexandrescu}},
  \bibnamefont{and} \bibinfo{author}{\bibfnamefont{L.}~\bibnamefont{Torner}},
  \bibinfo{journal}{Phys. Rev. A} \textbf{\bibinfo{volume}{68}},
  \bibinfo{pages}{050301} (\bibinfo{year}{2003}).

\bibitem[{\citenamefont{Wootters}(1998)}]{wootters:prl-80-2245}
\bibinfo{author}{\bibfnamefont{W.~K.} \bibnamefont{Wootters}},
\bibinfo{journal}{Phys. Rev. Lett.} \textbf{\bibinfo{volume}{80}},
\bibinfo{pages}{2245} (\bibinfo{year}{1998}); $T(\rho) =
[\max\{0,\lambda_1-\lambda_2-\lambda_3-\lambda_4\}]^2$, $\lambda_i$ are the
square roots of the eigenvalues of
$\rho(\sigma_2\otimes\sigma_2)\rho^*(\sigma_2\otimes\sigma_2)$ in
nonincreasing order by magnitude, with
$\sigma_2=\genfrac(){0cm}{1}{0\;-i}{i\;\;\;0}$.

\bibitem[{\citenamefont{James et~al.}(2001)\citenamefont{James, Kwiat, Munro,
  and White}}]{james:pra-64-052312}
\bibinfo{author}{\bibfnamefont{D.~F.~V.} \bibnamefont{James}},
  \bibinfo{author}{\bibfnamefont{P.~G.} \bibnamefont{Kwiat}},
  \bibinfo{author}{\bibfnamefont{W.~J.} \bibnamefont{Munro}}, \bibnamefont{and}
  \bibinfo{author}{\bibfnamefont{A.~G.} \bibnamefont{White}},
  \bibinfo{journal}{Phys. Rev. A} \textbf{\bibinfo{volume}{64}},
  \bibinfo{pages}{052312} (\bibinfo{year}{2001}).

\bibitem{binary}Binary plane-wave phase gratings~\cite{allen:2003-oam}
($\sim\!40\%$ diffraction efficiency) project the states $|g\rangle$,
$|l\rangle$, $|r\rangle$ and $\cos(\theta)|h\rangle+\sin(\theta)|v\rangle =
|l\rangle + e^{i2\theta} |r\rangle$ with $\theta = n\pi/8, n=-1,0,\ldots,8$
By displacing the holograms for $|l\rangle$ ($|r\rangle$) we project
arbitrary linear combinations~\cite{mair:nature-412-313} of $|g\rangle$ and
$|l\rangle$ ($|r\rangle$).

\bibitem[{\citenamefont{Clauser et~al.}(1969)\citenamefont{Clauser, Horne,
  Shimony, and Holt}}]{clauser:prl-23-880}
\bibinfo{author}{\bibfnamefont{J.}~\bibnamefont{Clauser}},
  \bibinfo{author}{\bibfnamefont{M.~A.} \bibnamefont{Horne}},
  \bibinfo{author}{\bibfnamefont{A.}~\bibnamefont{Shimony}}, \bibnamefont{and}
  \bibinfo{author}{\bibfnamefont{R.~A.} \bibnamefont{Holt}},
  \bibinfo{journal}{Phys. Rev. Lett.} \textbf{\bibinfo{volume}{23}},
  \bibinfo{pages}{880} (\bibinfo{year}{1969}).

\bibitem[{\citenamefont{Terhal}(2000)}]{terhal:pla-271-319}
\bibinfo{author}{\bibfnamefont{B.~M.} \bibnamefont{Terhal}},
  \bibinfo{journal}{Phys. Lett. A} \textbf{\bibinfo{volume}{271}},
  \bibinfo{pages}{319} (\bibinfo{year}{2000}).

\bibitem{leakage} Displaced plane-wave holograms allow small leakage of
unwanted states into the fiber~\cite{langford:prl-93-053601}.  This
potentially explains the smaller-than-predicted Bell parameter for the
non-maximally entangled spatial-mode states (e.g., $S_\text{exp}=2.28(1)$
versus the prediction $S_\text{pred}=2.35$).

\bibitem{tomodetails} Data in Fig.~2 (Fig.~3) were collected for $40\,$s
($20\,$s) per projection with $\sim\!600$ ($\sim\!100$) detected pairs/s.

\bibitem[{\citenamefont{Zyczkowski et~al.}(1998)\citenamefont{Zyczkowski,
  Horodecki, Sanpera, and Lewenstein}}]{zyczkowski:pra-58-883}
\bibinfo{author}{\bibfnamefont{K.}~\bibnamefont{Zyczkowski}},
  \bibinfo{author}{\bibfnamefont{P.}~\bibnamefont{Horodecki}},
  \bibinfo{author}{\bibfnamefont{A.}~\bibnamefont{Sanpera}}, \bibnamefont{and}
  \bibinfo{author}{\bibfnamefont{M.}~\bibnamefont{Lewenstein}},
  \bibinfo{journal}{Phys. Rev. A} \textbf{\bibinfo{volume}{58}},
  \bibinfo{pages}{883} (\bibinfo{year}{1998}).

\bibitem[{\citenamefont{White et~al.}(2001)\citenamefont{White, James, Munro,
  and Kwiat}}]{white:pra-65-012301}
\bibinfo{author}{\bibfnamefont{A.~G.} \bibnamefont{White}},
  \bibinfo{author}{\bibfnamefont{D.~F.~V.} \bibnamefont{James}},
  \bibinfo{author}{\bibfnamefont{W.~J.} \bibnamefont{Munro}}, \bibnamefont{and}
  \bibinfo{author}{\bibfnamefont{P.~G.} \bibnamefont{Kwiat}},
  \bibinfo{journal}{Phys. Rev. A} \textbf{\bibinfo{volume}{65}},
  \bibinfo{pages}{12301} (\bibinfo{year}{2001}).

\bibitem[{\citenamefont{Eisenberg et~al.}(2004)}]{eisenberg:prl-93-193901}
\bibinfo{author}{\bibfnamefont{H.~S.} \bibnamefont{Eisenberg}}
  \bibnamefont{et~al.}, \bibinfo{journal}{Phys. Rev. Lett.}
  \textbf{\bibinfo{volume}{93}}, \bibinfo{pages}{193901}
  (\bibinfo{year}{2004}).

\end{thebibliography}
\end{document}